\documentclass[11pt]{article}
\usepackage{amsfonts,amsmath,amsthm,array,amssymb}

\usepackage{graphicx}
\usepackage{picinpar,graphicx}
\usepackage{graphics}
\usepackage{multicol}
\usepackage[noblocks]{authblk}

\begin{document}

\title{Eternal Sunshine of the Solar Panel}
\author{Mackenzie Ginithan$^1$, Duber Gomez Fonseca$^2$, Daniel Lefevre$^3$, Sowmya Srinivasan$^4$, Barbara Urena$^5$, Kamal Barley$^5$, Jos\'e Vega$^5$, Kamuela E. Yong$^5$, Jos\'e Flores$^1$
 \\ \footnotesize{$^1$ University of South Dakota, $^2$ University of Houston Downtown, $^3$ Virginia Tech, $^4$ Bryn Mawr College, $^5$ Arizona State University}} 
 \date{\today}
\maketitle
\begin{abstract}
{\noindent The social dynamics of residential solar panel use within a theoretical population are studied using a compartmental model. In this study we consider three solar power options commonly available to consumers: the community block, leasing, and buying. In particular we are interested in studying how social influence affects the dynamics within these compartments. As a result of this research a threshold value is determined, beyond which solar panels persist in the population. In addition, as is standard in this type of study, we perform equilibrium analysis, as well as uncertainty and sensitivity analyses on the threshold value. We also perform uncertainty analysis on the population levels of each compartment.  The analysis shows that social influence plays an important role in the adoption of residential solar panels. }
\end{abstract}
\newpage
%
%
\section{Introduction}
The rapid growth of population in the United States has produced a sharp rise in the consumption of food, water, and electricity. Based on this increase in the population, governmental authorities are turning to renewable energy sources for solutions~\cite{tarp}. Additionally, private corporations have understood the necessity for clean renewable energy, and have invested time and capital into developing efficient methods of producing electricity from renewable sources, one of the most promising of which is solar energy. 

The first solid state solar cell, built in 1883 by Charles Edgar Fritts, successfully demonstrated that sunlight could be used as a viable energy source but was highly inefficient as less than one percent of the absorbed light was transformed into electric current~\cite{Cleveland2008}. The technology has advanced, and today photovoltaic solar cells are efficient sources of providing power~\cite{Epstein2011}. 

In this study we analyze the complex interactions between households that use solar power and those that do not, specifically at the effects of social influence on these interactions. With more households using solar technology, the population can become less reliant on coal and other ``unclean'' sources, thus decreasing the emissions of harmful chemicals that cause pollution.  We include a  framework for study of the significance of government subsidies and the subsequent effects on the system. Our goal is to examine how solar panel adoption is affected by social influence and how the technology can spread across a population over time. 


\section{Social Influence Model}

The spread of new technologies through populations and the importance of social influence to consumer decision making has been well-documented~\cite{powerofagoodidea},~\cite{popmodeling},~\cite{rogers05}. Furthermore, it has been shown that the amount of influence felt by the individual consumer grows in proportion to the size of the referent peer group ~\cite{Lascu1995}. It has also been reported that some consumers buy environmentally conscious products in order to increase social status~\cite{griskevicius}.  Marketers from different industries utilize socially conscious advertisements to capitalize on this behavior~\cite{griskevicius}. Since rooftop solar panels are a prominent feature on a house, their visibility intensifies the effects of peer interactions~\cite{bollinger}. Social influence has been similarly examined in other scenarios, such as predicting voting behavior~\cite{gerber} and the spread of mobile phone technology~\cite{kwon}. We aim to quantify and analyze the impact of this type of social influence on the spread of solar panel technology.

\begin{figure}[htb!]
\vspace{- 1 cm}
\begin{center}
\includegraphics[scale=0.7]{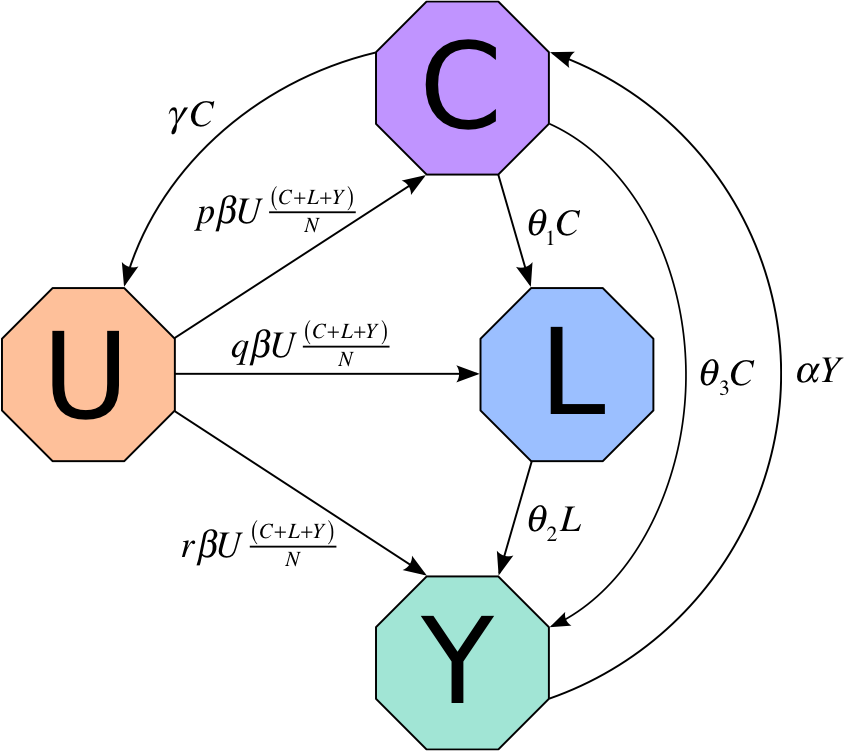}
\caption{\emph{Diagram describing interactions between non-solar power user households ($U$) and solar power user households ($C, L, Y$).}}
\label{fig:diagram}
\end{center}
\end{figure}

We develop a system of ordinary differential equations that describes the dynamics between electricity consuming households in a theoretical population, as shown in the compartmental model, see Figure~\ref{fig:diagram}. The population is compartmentalized by household according to their primary energy source into a non-solar power class $(U)$, solar power through a community block $(C)$, solar power through leasing $(L)$, and solar power through buying $(Y)$ solar panels. We assume that non-users of solar power (the $U$ class) receive their power from coal-fired generation, which currently accounts for more than half of the electricity produced in America~\cite{treehugger}. Utility companies operate externally located solar panel farms, from which they transport power to households of subscribers to the community block option. As such, for an extra monthly fee, the community block allows residents the use of clean energy without having to install a rooftop solar panel system. It also allows for households to opt for or out of the program and receive non-solar powered electricity with minimal effort~\cite{SRP2012}. The meanings of the state variables are summarized in Table~\ref{tab:state}. 

\begin{table}[htb!]
\caption{State Variables and Meanings}
\label{tab:state}
\begin{center}
\begin{tabular}{|c|p{10cm}|}
\hline
State Variable & Meaning\\ 
\hline
$U$ & Households not using any type of solar power\\
$C$ & Households using the community block option \\
$L$ & Households using solar panels they have leased\\
$Y$ & Households using solar panels they have bought \\
\hline
\end{tabular}
\end{center}
\end{table}

\begin{table}[h!] 
\caption{Parameter definitions for the model}
\label{tab:parameters}
\begin{center}
\begin{tabular}{|c| p{11cm}|}
\hline Parameter & Definition \\
\hline $\beta$ & Strength of social influence of solar power users per unit time\\
$\gamma$ & Per capita rate at which households move from community block to non-solar \\
$\theta_1$ & Per capita rate at which households move from community block to leasing \\
$\theta_2$ & Per capita rate at which households move from leasing to buying \\
$\theta_3$ &  Per capita rate at which households move from community block to buying \\
$\alpha$ & Per capita rate at which households move from buying to community block\\
$p,q,r$ & Proportion of households moving from non-solar to community block, leasing, buying, respectively\\
$u_0, c_0, l_0, y_0$ & Initial proportions of population in each class \\
 \hline
\end{tabular}
\end{center}
\end{table}

In the proposed model we assume that consumers receive power from only one source and not a combination of compartments. On an aggregate level, this assumption is valid if one considers that, for example, two households receiving 50\% solar power and 50\% coal power could be thought of as one household receiving 100\% solar power and one household receiving 100\% coal power. We do not explicitly model differences in economic status or power consumption of consumers, but note that these differences may be reflected in the values of each parameter.

Social influence effectiveness per unit time is denoted by $\beta$. The parameters $p$, $q$, and $r$ represent the proportion of housheolds that move out of $U$ and into $C$, $L$, and $Y$, respectively. Of the households moving out of $U$, $p$ represents the proportion that adopt the community block option, $q$ represents the proportion that elect to lease solar panels, and $r=1-p-q$ stands for the proportion that decide to buy solar panels. Assuming homogeneous mixing, social influence of solar panel users is represented by $\frac{C}{N}+\frac{L}{N}+\frac{Y}{N}$, i.e. the influence is dependent on the level of interaction rather than the number of solar panel users with whom households interact~\cite{carlos}. We assume also that each of the solar power user classes equal social pressure on households in $U$.

Since households in the community block can subscribe and unsubscribe to this option with relative ease, the model includes a flow from $C$ to $U$ at a rate $\gamma$, as shown in Figure~\ref{fig:diagram}. Because leasing and buying typically require significant investment of money and a long-term commitment ~\cite{SRP2012}, we assume that the amount of households who move from $L$ to $U$ or from $Y$ to $U$ is negligible, therefore we do not consider them in the model. 

We represent the rate of households switching from $C$ to $L$ as $\theta_1$, the rate from $L$ to $Y$ as $\theta_2$, and from $C$ to $Y$ as $\theta_3$. Consumers in the $L$ class sign a contract for a fixed period of time, commonly around 15 years~\cite{leasing}. We assume households will want to avoid penalties associated with breaking the lease agreement, so the flow from $L$ to $C$ is negligible, therefore we do not consider this flow. Lessees may, however, buy out their lease to move into the $Y$ class. Consumers in the $Y$ class own their own solar panels and have very little incentive to enter a lease agreement or stop using solar power entirely. However we assume that a small portion of this population may, for financial reasons, prefer to sell the panels and return to the $C$  class at a rate $\alpha$. Given our assumptions we consider a time scale of 30 years. 

The model reflecting the above dynamics is described as follows: 
\begin{equation}
\label{4eqns}
\begin{aligned}
U' &= -\beta U\left(\frac{C}{N}+\frac{L}{N}+\frac{Y}{N}\right)+\gamma C\\
C'& =p\beta U\left(\frac{C}{N}+\frac{L}{N}+\frac{Y}{N}\right)-\gamma C-\theta_1 C-\theta_3 C+\alpha Y\\
L'&=q \beta U\left(\frac{C}{N}+\frac{L}{N}+\frac{Y}{N}\right)+\theta_1 C-\theta_2 L\\
Y' &=r\beta U\left(\frac{C}{N}+\frac{L}{N}+\frac{Y}{N}\right)+\theta_2 L+\theta_3 C-\alpha Y
\end{aligned}
\end{equation}
where $N=U+C+L+Y$ represents the size of the total population, $r=1-p-q$, and the parameter definitions for the model are given in Table~\ref{tab:parameters}. We have made the simplifying assumption that the total population is constant in time and that the population is closed, thus 
$$\frac{dN}{dt}=0.$$ 
The constant and closed population allows us, for simplicity of analysis, to reduce System~\ref{4eqns} to three equations using the substitution $U=N-C-L-Y$. Scaling the population with the following substitutions: $u=\frac{U}{N}, c=\frac{C}{N}, l=\frac{L}{N}, y=\frac{Y}{N}$, where $u,c,l,y$ represent the proportion of the total population that are in each of the four classes, we obtain the following equivalent reduced system, on which we do our analysis: 
\begin{eqnarray}
\begin{aligned}
c'&=p\beta (1-c-l-y)(c+l+y)-\gamma c-\theta_1 c-\theta_3 c+\alpha y\\
l'&= q\beta(1-c-l-y)(c+l+y)+\theta_1 c-\theta_2 l \\
y'&=(1-p-q)\beta (1-c-l-y)(c+l+y)+\theta_2 l+\theta_3 c-\alpha y
\label{eqn:reduced}
\end{aligned}
\end{eqnarray}


\section{Stability Analysis}
System \ref{eqn:reduced} has two equilibria denoted by $E_0$, and $E_1$:
\begin{eqnarray*}
E_0 &=& (0,0,0)\\
E_1 &=& (c^*,l^*,y^*).
\end{eqnarray*}

\subsection{Non-Solar Equilibrium, $E_0$}
When the system is at $E_0$, no households are using solar power, and thus $u=1$. Using the next generation operator approach we compute the following threshold value $\mathfrak{T}$ ~\cite{carlos},~\cite{ngo}:
$$\mathfrak{T}=\frac{\beta[\alpha(q\gamma+\theta_1)+\theta_2(\alpha+\gamma(1-p)+\theta_1+\theta_3)]}{\alpha \gamma \theta_2}. $$
A detailed procedure of computing $\mathfrak{T}$ is given in Appendix A. Beyond this threshold value, solar power becomes an established source of residential power use, i.e. $\mathfrak{T}<1$ implies the amount of households using solar panels will taper off over time, but $\mathfrak{T}>1$ implies some nonzero proportion of households using solar power is sustained indefinitely in the population. Intuitively, we claim that when the threshold is crossed, the social pressure of solar panel users is strong enough to influence non-user households to adopt solar technology. 

The threshold $\mathfrak{T}$ can be rewritten as $$\mathfrak{T}=q\beta \left( \frac{1}{\theta_2}+\frac{1}{\alpha}\right)+r\beta\frac{1}{\alpha}+\frac{\beta}{\gamma}\left(1+\frac{\theta_1}{\theta_2}+\frac{1}{\alpha}(\theta_1+\theta_3)\right).$$
Note that: 
\begin{eqnarray*}
\beta \left(\frac{q}{\theta_2}+\frac{\theta_1}{\gamma \theta_2}+\frac{1}{\gamma}+\frac{1-p}{\alpha}+\frac{\theta_1+\theta_3}{\alpha \gamma}\right) \geq \beta \left(\frac{1}{\gamma}\right).
\end{eqnarray*}

In this form, the threshold is grouped into three terms, and its dependence on the model parameters is more directly observable. The first term represents the rate of flow into the leasing class from the non-solar class, and the flow from leasing through the buying class and back into the community class. The second term represents the rate of flow into the buying class from the non-solar class, and from the buying class back into the community class. The third term encompasses both the rate of flow out of the solar classes into the non-solar class, as well as the rates of exchange between the three solar classes. Since all of these terms are positive, if only one of them is greater than one, the entire threshold is greater than one and the population will tend towards the solar classes. We can interpret this to mean that if the flow into any one of the solar classes is greater than the flow out, the use of solar panels will persist in the population.

\subsection{Solar Equilibrium $E_1$}
At $E_2= (c^*,l^*,y^*)$, solar technology persists in the population at some nonzero level, where: 
\begin{eqnarray*}
c^*&=&\frac{\beta(\mathfrak{T}-1)}{\gamma \mathfrak{T}^2}\\
\vspace{1 cm}
l^*&=&\frac{\beta(\mathfrak{T}-1)}{(\gamma \theta_2 \mathfrak{T})^2}(q\gamma +\theta_1)\\
\vspace{1 cm}
y^*&=&\frac{\beta (\mathfrak{T}-1)}{(\alpha \gamma \mathfrak{T})^2}((1-p)\gamma +\theta_1+\theta_3).
\end{eqnarray*}
A sufficient condition for the existence of $E_1$ is $\beta>\gamma$. Using the Routh-Hurwitz criteria, we are able to determine sufficient conditions for the stability of $E_1$ ~\cite{carlos}, shown in Appendix B. When the system is at this equilibrium, solar power becomes established in the population over the course of our time scale. We attempted to find necessary conditions for the stability of $E_1$ but found that it offered us no practical interpretation. 


\section{Parameter Estimation}\label{sec:params}
\begin{figure}[htb!]
\begin{center}
\includegraphics[scale=0.7]{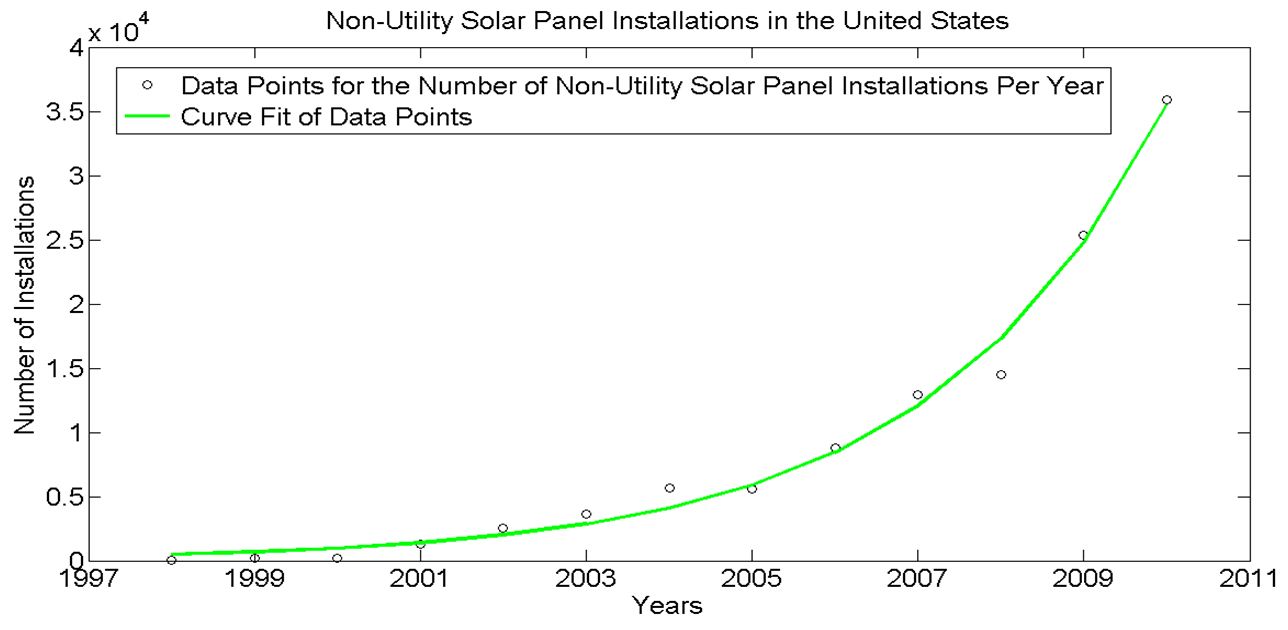}
\vspace{- 1 cm}
\caption{\emph{Grid-connected non-utility solar panel installations over time, and its fitted curve to estimate $\beta$.}}
\label{fig:data}
\end{center}
\end{figure}

In order to estimate the strength of social influence $\beta$, we use least squares regression estimation and estimate a 95\% confidence interval. The data used to estimate the parameter $\beta$, shown in Figure~\ref{fig:data} is taken from Table 1 of  Tracking the Sun~\cite{Barbose}. This data documents the number of grid-connected non-utility solar panel installations from 1998 to 2010, which we use as a proxy for the total number of households using solar panels, i.e. $C+L+Y$. The estimated parameter value for $\beta$ is $0.35981$ with a $95\% $ confidence interval of $[0.22317, 0.49645]$.
Equilibrium analysis of the system shows the threshold $\mathfrak{T}$ to be  $$\mathfrak{T}=\frac{\beta[\alpha(q\gamma+\theta_1)+\theta_2(\alpha+\gamma(1-p)+\theta_1+\theta_3)]}{\alpha \gamma \theta_2}.$$ Above this threshold, the use of solar panels in the population is sustained over time. As shown in Figure~\ref{fig:data}, the growth of solar panel usage is positive, implying $\mathfrak{T}>1$. 

Households move from $Y$ to $C$  at a rate $\alpha$. This movement is primarily made up of households who sell their solar panels and adopt the comunity block option for financial reasons. Associated with such movement are the large time costs of selling used solar panels; therefore $\alpha$ is the smallest parameter in the model. Since unsubscribing from the community block is a relatively easy process, $\gamma$, the rate of flow from $C$ to $U$, is the largest remaining parameter. Buying out a lease agreement is simpler than entering a lease or buying panels for the first time, so we assume $\theta_2>\theta_1$ and $\theta_2>\theta_3$. Since households who lease solar panels do not receive incentives~\cite{SRP2012}, we assume that households with the means would rather buy their own panels than enter a lease agreement, so $\theta_3 \geq \theta_1$. This reasoning yields the relationship $\beta \geq \gamma \geq \theta_2 \geq \theta_3 \geq \theta_1 \geq \alpha$. 


\section{Numerical Analysis}
\begin{figure}[h]
\begin{center}
\includegraphics[scale=0.6]{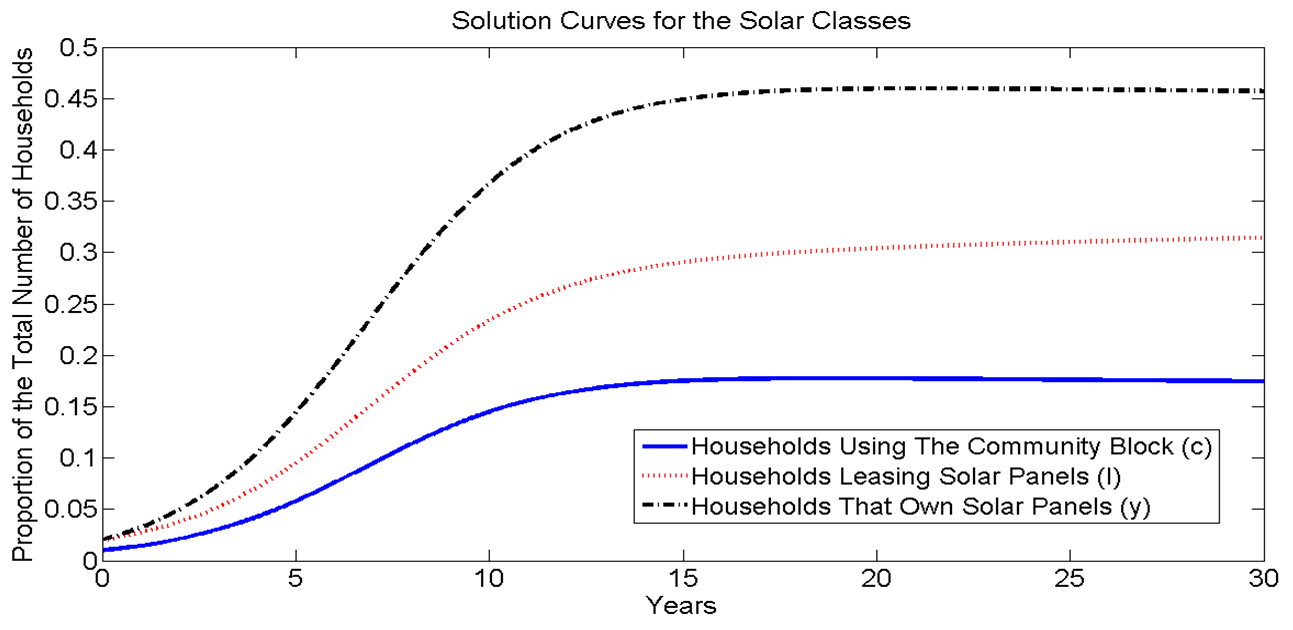}
\caption{\emph{Solution curves using $\beta$ value from estimated range. All three solar classes persist over time at a nonzero level.}}
\label{fig:slncurves1}
\end{center}
\end{figure}

\begin{figure}[h]
\begin{center}
\includegraphics[scale=0.6]{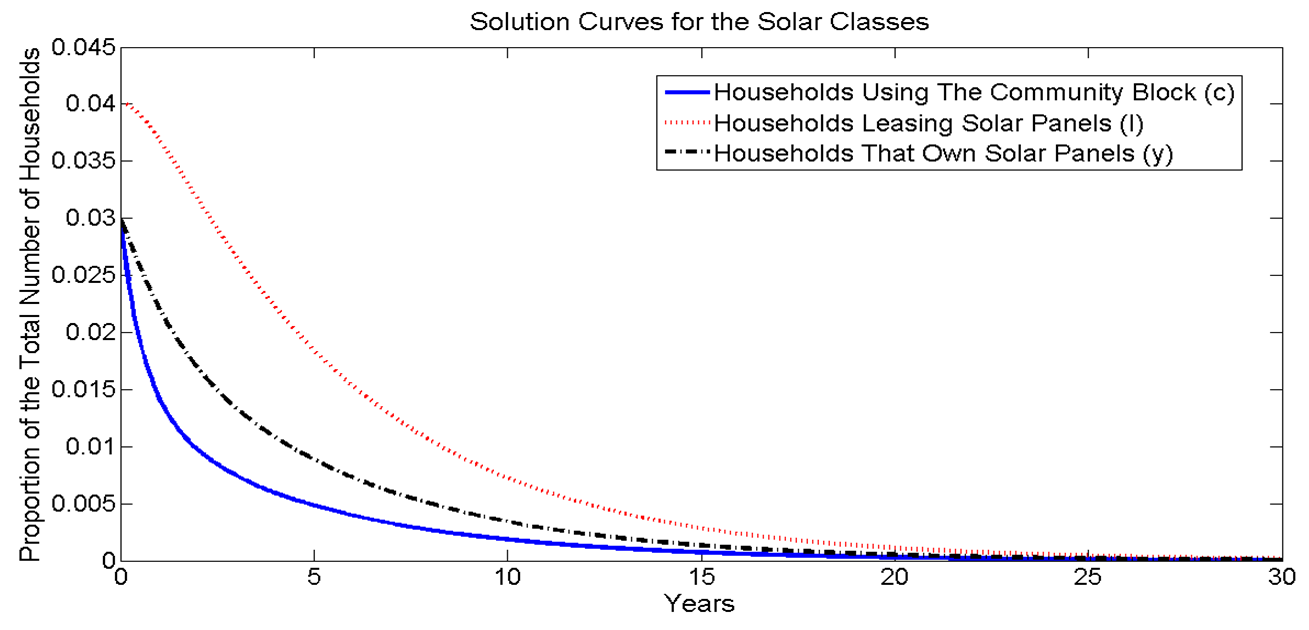}
\caption{\emph{Soution curves using $\beta$ value outside of estimated range. All solar classes tend toward zero and are not sustained over time.}}
\label{fig:slncurves2}
\end{center}
\end{figure}

Figures~\ref{fig:slncurves1} and~\ref{fig:slncurves2} show solution trajectories for the model plotted with the initial conditions and parameter values displayed in Table~\ref{tab:initialcon}. 
\begin{table}[htb!] 
\caption{Solution Curves with Parameters and Initial Conditions for Solution Curves}
\label{tab:initialcon}
\begin{center}
\begin{tabular}{|c|c|c|}
\hline Parameters& Figure~\ref{fig:slncurves1}& Figure~\ref{fig:slncurves2} \\ \hline
$\beta$ & 0.22317 & 0.01 \\ \hline
$\gamma$ & 0.065 & 0.7 \\ \hline
$\theta_1$ & 0.0006 & 0.2 \\ \hline
$\theta_2$ & 0.01 & 0.15 \\ \hline
$\theta_3$ & 0.0006 & 0.15 \\ \hline
$\alpha$ & 0.02 & 0.5\\ \hline
p & 0.2 & 0.5 \\ \hline
q & 0.3 & 0.3 \\ \hline
r & 0.5 & 0.2 \\ \hline
$c_0$ & 0.01 & 0.03 \\ \hline
$l_0$ & 0.02 & 0.04 \\ \hline
$y_0$ & 0.02 & 0.03 \\ \hline
\end{tabular}
\end{center}
\end{table}

Figure~\ref{fig:slncurves1} shows the solar classes approaching positive values of the proportion of the total number of households. In this figure the households using the community block ($C$ class) approach 17.46\% of the total population. The households leasing solar panels ($L$ class) approach 31.44\%, and the households buying solar panels ($Y$ class) approach 45.7\%. Given the initial conditions and parameters of Table~\ref{tab:initialcon}, System~\ref{eqn:reduced} approaches a positive equilibrium and solar power persists over time, with approximately 5\% of households using non-solar power. These conditions are favorable to the establishment of solar power as a central source of energy. 

Figure~\ref{fig:slncurves2} shows that the solar classes approach zero, while implying that the households not currently using solar power ($U$ class) approach 100\%, suggesting that over time, solar power cannot be established and that most of the population uses non-solar power in the long term. We note that this behavior is not necessarily realistic, since the behavior shown in Figure~\ref{fig:slncurves2} is not supported by the data in Figure~\ref{fig:data}. This plot in Figure~\ref{fig:slncurves2} is included in order to show the full range of behavior of the system. 
%

\section{Uncertainty \& Sensitivity Analysis on $\mathfrak{T}$}
\begin{table}[htb!] 
\title{Uncertainty Cases With $\mathfrak{T}$ Distribution Information}
\vspace{-.7 cm}
\begin{center}
\begin{tabular}{|l|l|l|l|l|}
\hline Cases & $\beta$ Range & $\mu$ of $\mathfrak{T}$ Distribution & $\sigma$ of $\mathfrak{T}$ Distribution & $\%$ $\mathfrak{T}$ Distribution $>$ 1 \\ \hline
1 & [0, 0.22317] & 126.2471 & 12707.8047 & $92.623\%$ \\ \hline
2 & [0.22317, 0.49645] & 442.5230 & 49028.7443 & $100\%$ \\ \hline
3 & [0.49645, 1] & 792.4298 & 71910.1342 & $100\%$ \\ \hline
\end{tabular}
\caption{ The parameter ranges for all cases are as follows: $\gamma \in[0, 0.3]$ , $\theta_1\in [0, 0.13]$, $\theta_2\in [0, 0.2]$ , $\theta_3 \in [0, 0.15]$, $\alpha \in [0, 0.1]$ , $p = 0.2, q = 0.3, r=0.5$. }
\label{tab:uncertainty}
\end{center}
\end{table}

Given the unavailability of comprehensive data that can be applied to our model, there is uncertainty associated with the estimation of the parameters. We perform uncertainty analysis on the threshold  $\mathfrak{T}$ in order to measure the variablilty in the threshold value that is caused by the uncertainty in parameter estimation. As part of the uncertainty analysis we consider each of the parameters $\beta, \theta_1, \theta_2, \theta_3, \alpha, \gamma$ as random variables with some probability density function, from which we then construct a frequency distribution for $\mathfrak{T}$ ~\cite{sallyblower}. In order to understand the effects of the uncertainty on the full system, in particular, the $U$ class, we perform uncertainty analysis on the system represented by System~\ref{4eqns}.

We use a sample size of 100,000 in our uncertainty analysis. Table~\ref{tab:uncertainty} describes three possible cases for parameter ranges. We assume a uniform distribution for each of the parameters and assume that $\alpha$, $\gamma$, $\theta_1$, $\theta_2$ and $\theta_3$ are constant throughout all three cases while varying the range of $\beta$. The ordering of parameters $\beta \geq \gamma \geq \theta_2 \geq \theta_1 \geq \theta_3 \geq \alpha$ remains the same as described in Section~\ref{sec:params} and we assume that the true values are contained in the intervals specified in Table~\ref{tab:uncertainty}.

Case 2 of Table~\ref{tab:uncertainty} corresponds to the $\beta$ range estimated in Section 4. Case 1 samples from below this estimated range, and Case 3 samples from above the estimated range. The results of the uncertainty analysis for all three cases show that the threshold $\mathfrak{T>1}$ for a large percentage, 93-100\%, of the $\mathfrak{T}$ distribution. These results show that given the conditions assumed in our model, solar power is likely to become an established source of energy over time. 

\begin{table}[htb!]
\caption{Sensitvity index for each parameter}
\label{tab:sensitivity}
\begin{center}
\begin{tabular}{|c|c|}
\hline Sensitvity Index & Value \\
\hline $S_\beta$ & $1-\frac{\theta_2(\alpha+\gamma(1-p)+\theta_1+\theta_3)}{(\alpha(q\gamma+\theta_1)+\theta_2}$\\
$S_\alpha$ &$ \frac{\theta_2 ((p-1)\gamma-\theta_1 -\theta_3)}{q\alpha \gamma +\theta_1 (\alpha+\theta_2)+\theta_2(\alpha+\gamma(1-p)+\theta_3)}$ \\
$S_\gamma$ & $-\frac{(\theta_1(\alpha+\theta_2)+\theta_2(\alpha+\theta_3)}{q\alpha \gamma+\theta_1(\alpha+\theta_2)+\theta_2(\alpha+\gamma-p\gamma+\theta_3}$\\
$S_{\theta_2}$ &$-\frac{\alpha(q\gamma+\theta_1)}{q\alpha \gamma+\theta_1(\alpha+\theta_2)+\theta_2(\alpha+\gamma-p\gamma+\theta_3}$\\
$S_{\theta_1} $& $\frac{\theta_1(\alpha+\theta_2)}{q\alpha \gamma+\theta_1(\alpha+\theta_2)+\theta_2(\alpha+\gamma-p\gamma+\theta_3}$\\
$S_{\theta_3} $& $-\frac{\theta_2 \theta_3}{q\alpha \gamma+\theta_1(\alpha+\theta_2)+\theta_2(\alpha+\gamma-p\gamma+\theta_3}$\\ \hline 
\end{tabular}
\end{center}
\end{table}

Local sensitivity analysis on $\mathfrak{T}$ allows us to measure how sensitive the threshold value is to small changes in its input parameters~\cite{mtbi04},~\cite{sallyblower}. We construct a sensitivity index for each of the parameters that determine $\mathfrak{T}$, shown in Table~\ref{tab:sensitivity}. For each parameter $\rho$, the sensitivity index $S_\rho$, which represents the normalized change in $\mathfrak{T}$ caused by a small change in $\rho$~\cite{mtbi04}, is given by :
$$S_\rho=\frac{\rho}{\mathfrak{T}}\frac{\partial \mathfrak{T}}{\partial \rho}.$$
 

\section{Uncertainty Analysis on Population Sizes Over Time}
\begin{figure}[h]
\begin{center}
\includegraphics[scale=0.7]{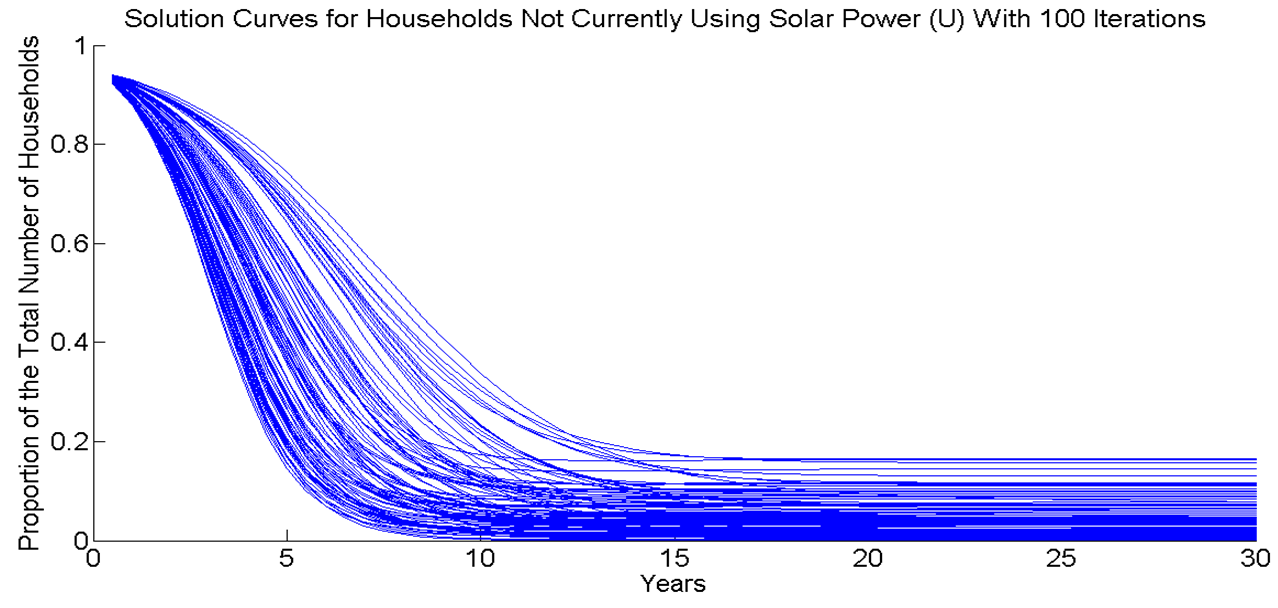}
\vspace{-1 cm}
\caption{\emph{Solution curves for the $U$ class for 100 iterations.}}
\label{fig:spaghettiu}
\end{center}
\end{figure}
\begin{figure}[h]
\begin{center}
\includegraphics[scale=0.7]{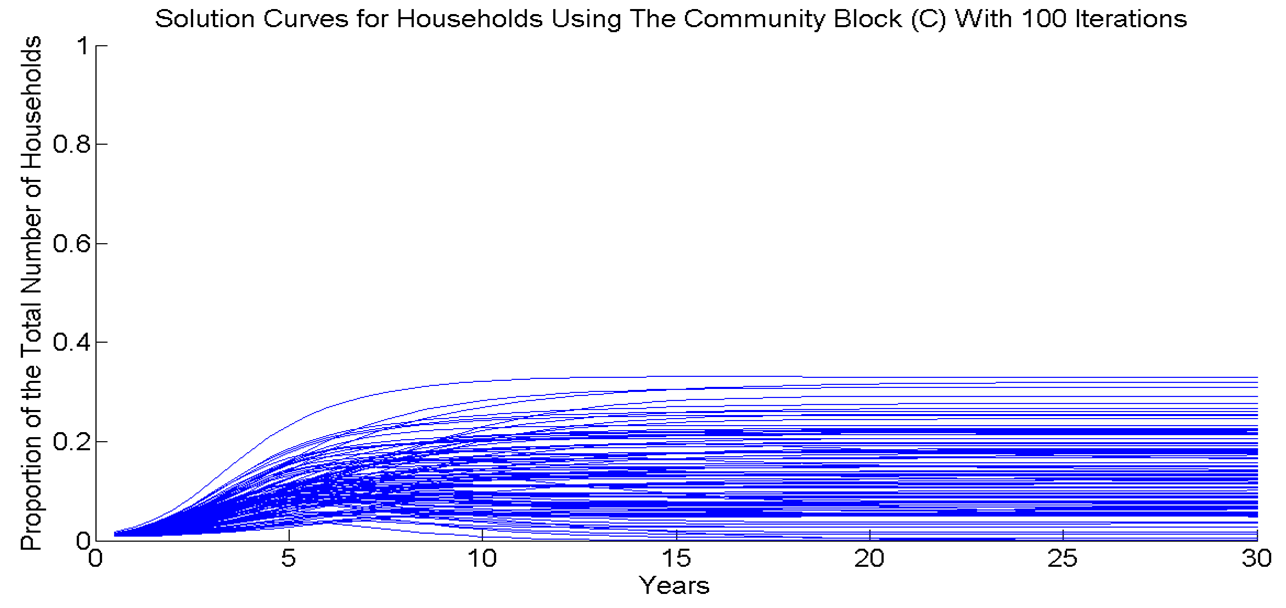}
\vspace{-1 cm}
\caption{\emph{Solution curves for the $C$ class for 100.}}
\label{fig:spaghettic}
\end{center}
\end{figure}
\begin{figure}[h]
\begin{center}
\includegraphics[scale=0.7]{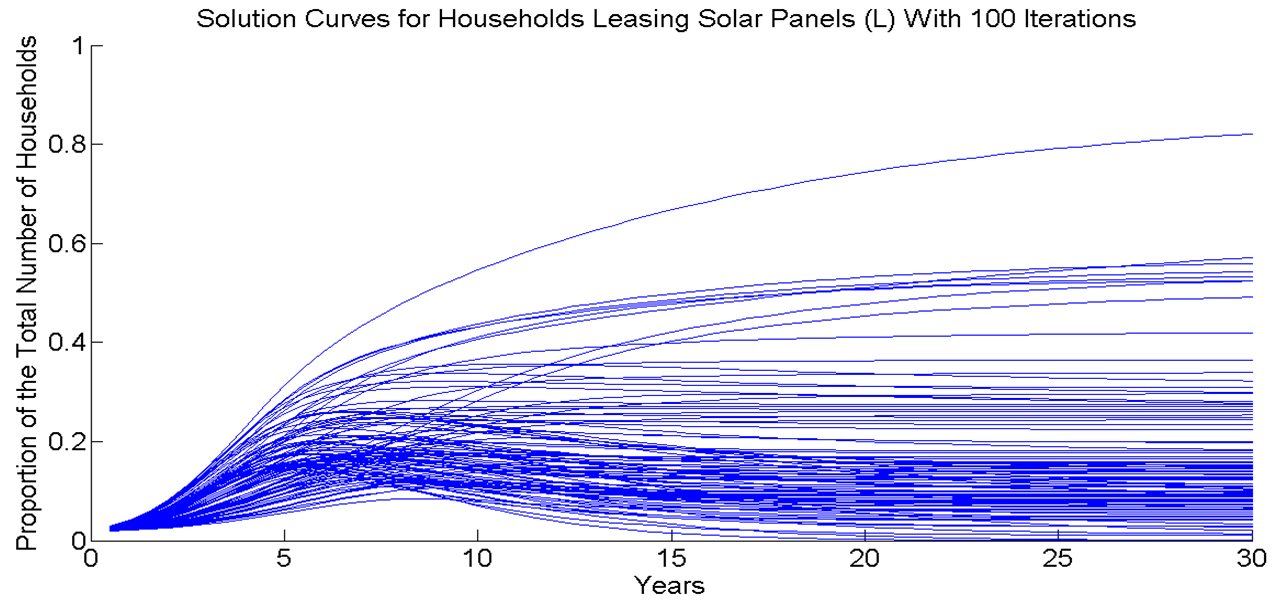}
\vspace{-1 cm}
\caption{\emph{Solution curves for the $L$ class for 100 iterations with dashed line indicating average solution curve.}}
\label{fig:spaghettil}
\end{center}
\end{figure}
\begin{figure}[ht]
\begin{center}
\includegraphics[scale=0.7]{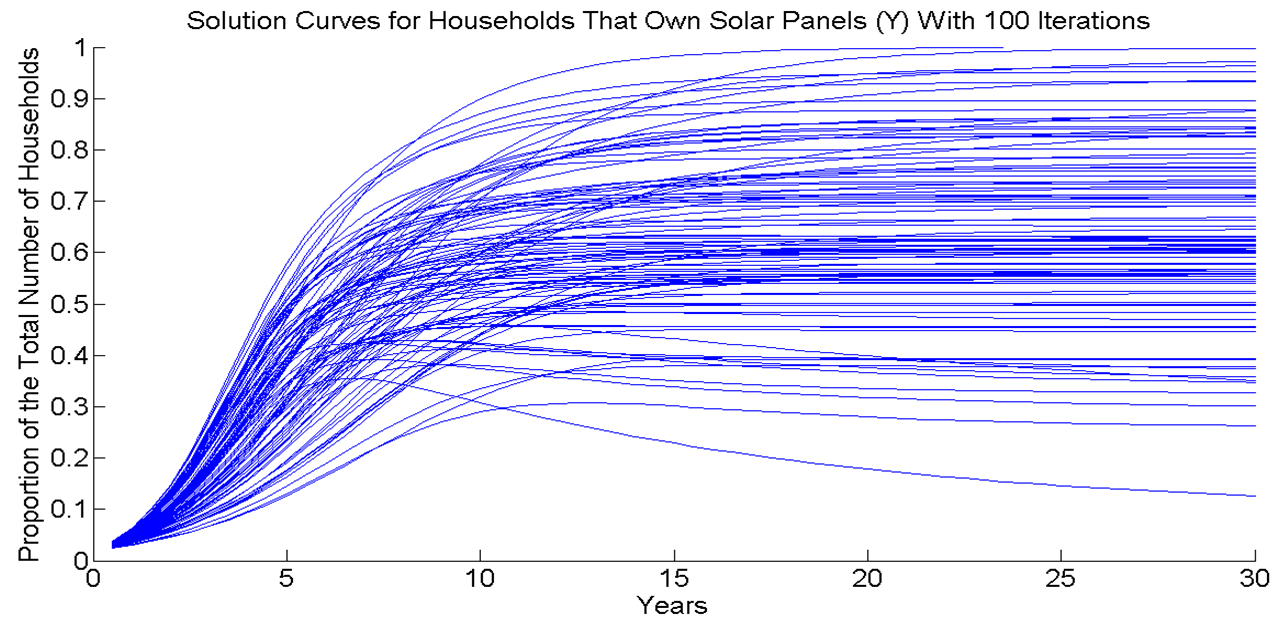}
\vspace{-1 cm}
\caption{\emph{Solution curves for the $Y$ class for 100 iterations with dashed line indicating average solution curve.}}
\label{fig:spaghettiy}
\end{center}
\end{figure}

There is uncertainty associated with the estimation of our parameters. We use a uniform distribution to generate random parameter values and plot solution curves for each set of values. The results can be seen in Figures~\ref{fig:spaghettiu}-~\ref{fig:spaghettiy}. The $U$ class shows the least amount of variation over time. The point at which its distribution is the widest is around the seven year mark. After that point the solution curves are grouped within the $[0,0.2]$ interval. The relatively low variation among the 100 iterations of solution curves indicates that the estimates of the $U$ class are fairly reliable given the uncertainty in the parameter estimation.

The solar classes have relatively wide distributions over time, with most of their curves tending toward nonzero levels. Figures~\ref{fig:spaghettiu}-~\ref{fig:spaghettiy} indicate that variation in the model's parameters has a significant influence on the final distribution among the solar classes, while the non-solar class tends to  exhibit similar behavior even for various sets of parameter values. We note also that there is greater variation in the final distributions of $L$ and $Y$ when compared with the variation in the distributions of $U$ and $C$. This indicates that given the uncertainty in our parameter estimates, the estimates of the size of the non-user and community block classes are likely to be more reliable than the estimates of the leasing or buying classes. We may conclude that accurate estimation of the relevant parameters is of critical importance if this model is to be applied to the design of policy.

\section{Discussion}
The widespread use of solar power has significant implications for society at large. The current electrical grid, fragile and overburdened, will benefit tremendously from increased use of solar power. Widespread power outages currently represent a significant cause of economic loss~\cite{costofoutage}. Since at least some of their energy consumption is covered by a localized source, households with their own solar panels are less reliant on the centralized grid system, so they are not as harshly affected by power outages~\cite{powerfromsun}. 

The benefits of solar power will extend to those who do not use it. Households can send the excess power generated by their solar panels back to the utility company, thus reducing the load on the power grid. This also helps in reducing pollution produced from using other power sources such as coal, which currently provides over half of the electricity consumed in the United States~\cite{treehugger}.  Coal produces high levels of pollution, which has associated economic costs~\cite{Epstein2011}. In addition to environmental harm, medical conditions such as lung cancer and heart attacks, which are leading causes of death in America, have been linked to coal pollution~\cite{Epstein2011}. Solar power is a clean energy source that causes relatively little environmental harm in its production or use~\cite{Weisser}. Increased usage of solar power could therefore contribute to a healthier population and possibly even lower death rates in areas where coal is a primary source of power.

We have considered a model in which the environmental concerns of the early adopters and social pressure are the only relevant factors prompting individuals to switch to solar power. It seems logical that any capital put into ad campaigns or incentives would serve to compound the effects of social influence.  For very small values of $\beta$, there may be a situation in which the threshold is not surpassed, but could be overcome with added effort from authorities and utility providers. After a certain level of the population has adopted the technology, social pressure alone may be enough to sustain the trend.  This observation is important to the authorities responsible for making decisions regarding our energy sources. In our simulations, we observed the population of the non-solar class always tended to a small fraction of the total population, but was still subject to a certain amount of variation. Therefore, future study could examine the amount of money spent on maximizing the number of households that adopt solar technology.

\newpage
\noindent \textbf{Acknowledgments} \\
We would like to thank Dr.~Carlos Castillo-Ch\'avez, Executive Director of the Mathematical and Theoretical Biology Institute (MTBI), for giving us the opportunity to participate in this research program.  We would also like to thank Co-Executive Summer Directors Dr.~Erika T.~Camacho and Dr.~Stephen Wirkus for their efforts in planning and executing the day to day activities of MTBI, and Preston Swan for his undying commitment to MTBI. We would like to thank our lead advisor, Dr. Jos\'e Flores for his cheerful attitude and Mathematica wizardry. We are especially grateful to Dr. Anuj Mubayi for his immense support in directing our ideas; Steve and Erika for their incredible patience and encouragement; Dr. Baojun Song for helping us pay attention to the details of our model; Dr. Juan Aparicio for providing simple but extremely insightful ideas; Kamal Barley for his amazing abiltiy to find journal articles related to any subject; Jos\'e Vega-Guzman and Dr. Kamuela Yong for their timely guidance and untiring revisions; Rachel Neu McCleary, Kailee Gray, and Dr. Catalin Georgescu for their unwavering support and enthusiasm for math; Dr. Fabio Sa\'nchez for his positivity, advice, and faith in our success all the way from Costa Rica; Dr. Sunmi Lee for her smile and lessons on optimal control; Emmanuel Morales, Romie Morales, and Diego Chowell for their help with MATLAB and general support; and all other MTBI participants, grad students, and faculty for helping us in any capacity. This research was conducted in MTBI at the Mathematical, Computational and Modeling Sciences Center (MCMSC) at Arizona State University (ASU). This project has been partially supported by grants from the National Science Foundation (NSF - Grant DMPS-0838705), the National Security Agency (NSA - Grant H98230-11-1-0211), the Office of the President of ASU, and the Office of the Provost of ASU.


\newpage
\section{Appendix}

\subsection{Appendix A: Computation of $\mathfrak{T}$}
Using the Next Generation Operator approach, we perform the following steps to calculate $\mathfrak{T}$: 
$$\mathfrak{[F]}= \begin{pmatrix} p\beta(1-c-l-y)(c+l+y) \\ q\beta(1-c-l-y)(c+l+y) \\ r\beta (1-c-l-y)(c+l+y)\end{pmatrix}  \mathrm{and} \hspace{.3 cm}
\mathfrak{[V]}=\begin{pmatrix} \gamma c+\theta_1c+\theta_3c-\alpha y \\ -\theta_1 c+\theta_2 l \\ -\theta_2 l+\alpha y-\theta_3 c \end{pmatrix}$$
From this we obtain: 
$$F=\begin{pmatrix}p\beta&p\beta&p\beta \\ q\beta&q\beta&q\beta \\ r\beta&r\beta&r\beta \end{pmatrix} \mathrm{and} \hspace{.3 cm}
V=\begin{pmatrix}\gamma+\theta_1+\theta_3&0&-\alpha \\ -\theta_1&\theta_2&0 \\ -\theta_3&-\theta_2&\alpha \end{pmatrix}$$
The product $FV^{-1}$ yields three eigenvalues, two of which are zero. The threshold value is the largest of these eigenvalues: \\
$$\mathfrak{T}=\frac{\beta[\alpha(q\gamma+\theta_1)+\theta_2(\alpha+\gamma(1-p)+\theta_1+\theta_3)]}{\alpha \gamma \theta_2}.$$

\subsection{Appendix B: Stability of $E_1$}
The characteristic polynomial of the Jacobian at $E_1$ is given by $\lambda^3+a_1 \lambda^2+a_2 \lambda+a_3$. Thus, $E_1$ is stable if the coefficient of the characteristic polynomial ($a_1, a_2 $ and $a_3$) satisfty the Routh-Hurwitz criteria with
$a_1>0$, $a_3>0$, and $a_1 a_2>a_3$.
\begin{equation*}
\begin{aligned}
a_1 &= \alpha + \beta + \gamma +\theta_1+ \theta_2 +\theta_3- \frac{2 \beta}{\mathfrak{T}}\\
a_2 &= \alpha (\beta-\gamma)+\beta \gamma(1-p)+ (\alpha+\beta)(\theta_1+\theta_2)+ \theta_2 (\gamma+\theta_1 +\theta_3) + \beta \theta_3 +\frac{2\beta (q \alpha \gamma+\alpha \theta_1- \theta_2^2)}{\theta_2 \mathfrak{T}}\\
a_3 &= \alpha \gamma \theta_2 (\mathfrak{T}-1) .
\end{aligned}
\end{equation*}
That is: 
\begin{multline*}
\begin{aligned}
& \left(\beta (\gamma(1-p)+\theta_1+\theta_2+\theta_3) +\alpha\left(\theta_1+\theta_2+\beta-\gamma \right) +\theta_1\theta_2 +\theta_2\theta_3 +\gamma \theta_2 +2\beta \left(\frac{q\alpha \gamma+\alpha \theta_1-\theta_2^2}{\theta_2 \mathfrak{T}} \right) \right) \\
&\left( \alpha+\beta+\gamma+\theta_1+\theta_2+\theta_3+\frac{2\beta}{\mathfrak{T}}\right) >\mathfrak{T}\alpha \theta_2 \gamma
\end{aligned}
\end{multline*}

\subsection{Appendix C: Model with Pollution Effects}
We attempted to study the effects of government subsidies and the corresponding response of pollution levels on the existence of solar panels over time. Amount of pollution $P$ is quantified by the following function:
$$P(U)=\frac{U}{2-U}. $$
The function is bounded between $0$ and $1$ where $0$ corresponds to the minimum amount of pollution and $1$ to the maximum of pollution. For $U=0$, $P(U)=0$, which implies that when there are no households using coal, the population produces no pollution. When $U=1$, $P(U)=1$, which suggests that when the entire population is using coal, pollution is at the highest possible level. Thus we can view $P$ as the amount of pollution that a population can produce, dependent on the proportion of households using coal.

The federal government, in an indirect effort to control pollution, provides subsidies and other incentives to households that buy solar panels. When the level of pollution changes, the government responds by changing the amount of subsidies accordingly. The population's reaction, depending on the direction of the subsidies change, is either for more households to buy solar panels or for less households to buy solar panels. This in turn affects the pollution levels. In order to incorporate this feedback into the model we make $\beta$, the rate at which households move into the solar panel classes, dependent on the amount of pollution: 
$$\beta(P)=\frac{P}{1+P}.$$
Note that since $P=\frac{U}{2-U}$, $\beta (P)=\frac{U}{2}$. The choice of this $\beta$ function is motivated by the need for a smooth positive increasing function whose first derivative decreases as $P$ approaches $1$. We need the first derivative to be decreasing since as the amount of pollution produced increases, we expect that the rate of households switching to solar would increase rapidly at first, and then level off as the amount of pollution reaches saturation.

We represent the amount of subsidies $S$ through the following, as a function of the buyer class $Y$: \\
$$S(Y)=s (1-p-q)\beta (P)U(C+L+Y)N,$$
where $s$ represents the average per household amount spent by the government on a subsidy, $(1-p-q)\beta U(C+L+Y)$ is the rate of people flowing from $U$ to $Y$, and $N$ is the total number of households in the population.

\subsection{Appendix D: Stochastic Model}
In this study we attempted to implement a stochastic version of our model with the Gillespie algorithm. Our motivation for this was to determine a projection for the likely amount of solar users over time given realistic conditions. When the stochastic model was created we found that the output did not give realistic results. We found that in the first time step a majority of non-solar panel households moved into solar classes. This behavior is not realistic since one time step represents one year.

Initially we thought that this behavior was a result of an error in our implementation of the Gillespie algorithm. But after close inspection we found that the rate of households moving from non-solar to solar classes completely dominated the other rates. Also, since we were using a relatively large total population, events occurred very frequently. The combination of these two conditions drive most households from non-solar to solar classes very fast. Therefore, we conclude that there is nothing wrong with our model, or our implementation of the Gillespie algorithm, but that the algorithm is inappropriate to use with our model.  For future work we suggest a method that overcomes these obstacles, perhaps a discrete time Markov chain in which the time steps are marked by one year.

\end{document}